# Solving the 3-SAT problem using network-based biocomputation

Jingyuan Zhu[1, 2, †], Aseem Salhotra[1,3, †], Christoph Robert Meinecke[4], Pradheebha Surendiran[1, 2], Roman Lyttleton[1, 2], Danny Reuter[4, 5], Hillel Kugler[6], Stefan Diez[7], Alf Månsson[1, 3,*], Heiner Linke[1, 2, *], Till Korten[7, *]

**Affiliations**
1. NanoLund, Lund University, Box 118, 22100 Lund, Sweden
2. Solid State Physics, Lund University, Box 118, 22100 Lund, Sweden
3. Department of Chemistry and Biomedical Sciences, Linnaeus University, 391 82 Kalmar, Sweden
4. Center for Microtechnologies, Technische Universität Chemnitz, Chemnitz, 09126, Germany
5. Fraunhofer Institute for Electronic Nano Systems ENAS, Chemnitz, Germany
6. Faculty of Engineering, Bar-Ilan University, Ramat Gan, Israel
7. B CUBE - Center for Molecular Bioengineering, Technische Universität Dresden, Dresden, 01307, Germany

†These authors contributed equally to this work
*Corresponding author: alf.mansson@lnu.se, heiner.linke@ftf.lth.se, till.korten@tu-dresden.de

**Abstract**
The 3-Satisfiability Problem (3-SAT) is a demanding combinatorial problem, of central importance among the non-deterministic polynomial (NP) complete problems, with applications in circuit design, artificial intelligence and logistics. Even with optimized algorithms, the solution space that needs to be explored grows exponentially with increasing size of 3-SAT instances. Thus, large 3-SAT instances require excessive amounts of energy to solve with serial electronic computers. Network-based biocomputation (NBC) is a multidisciplinary parallel computation approach with drastically reduced energy consumption. NBC uses biomolecular motors to propel cytoskeletal filaments through nanofabricated networks that encode the mathematical problems. By stochastically exploring possible paths through the networks, the cytoskeletal filaments find possible solutions to the encoded problem instance. Here we first report a novel algorithm that converts 3-SAT into NBC-compatible network format. We demonstrate that this algorithm works in practice, by experimentally solving four small 3-SAT instances (with up to 3 variables and 5 clauses) using the actin-myosin biomolecular motor system. This is a key step towards the broad general applicability of NBC because polynomial conversions to 3-SAT exist for a wide set of important NP-complete problems.

## Introduction

The Satisfiability Problem (SAT) is a combinatorial decision problem that evaluates Boolean (binary) logic (see Textbox 1). Solving SAT is essential for technologies such as symbolic model checking [1], planning in artificial intelligence [2], automated theorem proving [3], and hardware verification [4]. A specific representation of SAT, namely the three satisfiability (3-SAT) problem, is also increasingly recognised to be of scientific importance in its own right. For example, a recent study showed all Ising models, of fundamental importance in magnetism and beyond [5,6], to be equivalent to instances of 3-SAT [7].

From a mathematical perspective, 3-SAT is considered to be the benchmark nondeterministic polynomial-time complete (NP-complete) problem [8]. Additionally, this problem is of particular interest because polynomial conversions to SAT exist for a wide range of other important NP-complete problems [9]. Whereas 3-SAT is conceptually simple, it is computationally demanding to solve in practice. Even though extensive efforts have been expended to solve 3-SAT on serial transistor-based computers [10,11], only algorithms with exponential worst-case complexity have been developed [11]. Therefore, the processing time and the energy cost of solving a 3-SAT instance grow exponentially with problem size. The demand for more computational power, and the need for reducing the energy used in computation, therefore necessitate the search for energy-efficient alternatives to serial, electronic computers [12,13].

An emerging approach that has been estimated to be very energy-efficient [14] is network-based biocomputation (NBC). So far, this approach has been used to solve a small instance of the NP-complete Subset Sum Problem (SSP) [14,15], and an optimize algorithm has been developed for solving Exact Cover with NBC[16]. In NBC, a given combinatorial problem is encoded into a graphical, modular network that, after implementation by nanofabrication, is explored in a parallel manner by biological agents [15]. This novel computation approach is an inherently stochastic method based on non-deterministic operation due to the random choice that agents make in the network. Similar to quantum computing, the output of NBC is a sample of the population of all possible solutions [17]. The result of NBC is therefore reported as the solution of the problem instance together with a statistical confidence interval. [14,15].

Here we take a key step in demonstrating the general applicability of NBC: We first present a new network algorithm for encoding 3-SAT instances in a graphic, planar layout suitable for NBC. We then implement this algorithm to experimentally solve four instances of 3-SAT by allowing myosin-driven actin filaments to explore the nanofabricated network that encodes the problem. Finally, using a newly developed statistical approach, we find solutions that are accurate on a 99% confidence level demonstrating the feasibility of our approach towards generalization of NBC.

# Results
## 3-SAT network algorithm

3-SAT (Textbox 1) is a variant of SAT that is restricted to a Boolean formula in conjunctive normal form (CNF), where each clause is limited to at most three literals. In 3-SAT, one asks whether the formula is satisfiable by assigning appropriate logical values to its variables. In the present study, we map 3-SAT to a spatially encoded NBC network that solves the mathematical problem. The network encoding is a critical step in

> **Textbox 1: The 3-Satisfiability Problem (3-SAT)**
>
> A propositional logic formula, also called Boolean expression, is built from variables, operators AND (conjunction, also denoted by $\wedge$), OR (disjunction, $\vee$), NOT (negation, $\neg$), and parentheses. For example:
>
> $\phi = (x_1 \vee x_2 \vee x_3) \wedge (\neg x_1 \vee x_2) \wedge (x_1 \vee \neg x_2) \wedge (\neg x_1 \vee x_2)$
>
> A literal is either a variable, called positive literal (e.g. $x_1$), or the negation of a variable, called negative literal (e.g. $\neg x_1$).
>
> A clause is a collection of literals enclosed in parentheses (e.g. $(x_1 \vee x_2 \vee x_3)$).
>
> A formula is in conjunctive normal form (CNF) or clausal normal form if it is a conjunction of one or more clauses. In CNF all clauses are disjunctions (OR). Any propositional logic formula can be represented in CNF form.

implementing NBC for a given NP-complete problem. We encode 3-SAT into network format in five steps. Briefly (see below for a detailed explanation including examples): *Step 1*: Literal assignments are encoded as binary numbers reflecting specific clauses that are satisfied by the respective literals [3,18]. *Step 2:* The binary numbers are represented by unary encoding in a planar network. *Step 3:* The bitwise 'OR' operation is realized as an 'OR-block' in the unary network via several types of junctions. *Step 4:* The entire 3-SAT instance is represented in a network consisting of one 'OR-block' per variable. *Step 5:* The network is further optimized to reduce the physical size (Figure 2). Satisfiability of the instance is tested at the target exit representing TRUE assignment for all clauses via the readout of the number of exiting computing agents. Notably, the algorithm for network design (Steps 1-5) scales polynomially with network size and requires no prior information about the solution.

As a minimal example, we use the 3-SAT instance $\Phi_1$ with two variables ($x_1$ and $x_2$) and four clauses: $(x_1 \vee x_2)$, $(\neg x_1 \vee x_2)$, $(x_1 \vee \neg x_2)$ and $(\neg x_1 \vee x_2)$ (Table 1):

$$\phi_1 = (x_1 \vee x_2) \wedge (\neg x_1 \vee x_2) \wedge (x_1 \vee \neg x_2) \wedge (\neg x_1 \vee x_2) \quad (1)$$

$\Phi_1$ was chosen such that exactly one assignment ($x_1 = TRUE, x_2 = TRUE$) satisfies the logical statement. In addition, we also solve three other instances $\Phi_2$, (two variables, four clauses, no solution), $\Phi 3$ (three variables, five clauses, one solution), and $\Phi_4$ (three variables, five clauses, one solution):

$$\phi_2 = (x_1 \vee x_2) \wedge (\neg x_1 \vee x_2) \wedge (x_1 \vee \neg x_2) \wedge (\neg x_1 \vee \neg x_2) \quad (2)$$

$$\phi_3 = (x_1 \vee x_2) \wedge (\neg x_1 \vee x_2) \wedge (x_1 \vee \neg x_2) \wedge (\neg x_1 \vee \neg x_2 \vee x_3) \wedge (x_1 \vee x_2 \vee x_3) \quad (3)$$

$$\phi_4 = (x_1 \vee x_2) \wedge (\neg x_1 \vee x_2) \wedge (x_1 \vee \neg x_2) \wedge (\neg x_1 \vee x_2) \wedge (x_1 \vee x_2 \vee x_3) \quad (4)$$

We use $\Phi_1$, as an example to explicitly walk through the five steps of the network encoding algorithm:

Step 1: Binary representation of 3-SAT instance $\Phi_1$

In the conjunctive normal form representation of 3-SAT, any variable assignment will either fulfil a clause (the result of the clause is TRUE) or not. Thus, an instance with $c$ clauses and $n$ variables can be represented by $n$ pairs of $c$-digit binary numbers: each digit of the binary numbers represents a clause and each pair of binary numbers represents the two possible assignments for each variable [3]. The digit is binary $1_2$ if the variable assignment fulfils the clause and $0_2$ otherwise (the 2 in subscript indicates the binary base). For our minimal example $\Phi_1$, the TRUE assignment of $x_1$ (first row in Table 1), fulfils the first $(x_1 \vee x_2)$ and the third ), $(x_1 \vee \neg x_2)$ clauses, while the second $(\neg x_1 \vee x_2)$ and fourth $(\neg x_1 \vee x_2)$ clauses are not fulfilled. Therefore, the respective binary representation is $1010_2$. We give the complete binary representation for our minimal example $\Phi_1$ in Table 1.

The whole 3-SAT instance is satisfiable if a combination with exactly one binary number from each pair can be found, such that the bitwise 'OR' operation for the combination returns all ones, indicating that each clause has been fulfilled. Table 2 lists all possible combinations of binary representations for $\Phi_1$. The combination $(1010_2 \wedge 1101_2 = 1111_2)$ corresponding to $(x_1 = TRUE, x_2 = TRUE)$ satisfies $\Phi_1$, indicating that this instance is satisfiable.

Step 2: Encoding of numbers in the network

In our spatial network encoding method, information is processed in a unary numeral system where a symbol (in our case the network column) representing 1 is repeated $N$-times to represent number $N$. In the network, we define the leftmost column as column '0' so that any agent in that column will represent the number '0'. Moving to the right, the number for each column changes in increments of 1. For example, in Figure 1a, from left to right, each column represents a number from $0_{10} = 0000_2$ to $7_{10} = 0111_2$. An agent travelling in a given column thus represents the corresponding number.

Step 3: Bitwise OR operation in unary encoding

Operations within the network are performed by connecting an input (entrance) column to the respective output (exit) column. The network consists of a grid of individual junctions, where different types of junctions allow agents to travel either straight down – not changing the input – or diagonally down – changing the input by one for each row they travel (see Supplementary Table S2 for details). Thus, the number of network rows required for each operation is the largest possible difference between input and output. By repeating the junctions at each column, the network block always performs the same operation with the same operand for a certain range of input columns.

Figure 1b shows an illustrative example network that performs an OR operation with operand $0101_2$ for all inputs from $0000_2$ to $1000_2$. The number of rows in the network equals the value of the operand. Each input is connected to the respective output by first moving the respective number of rows straight down and then moving the respective number of rows diagonally down. Within the network, each input consists either of a *reset-FALSE junction* (see Table S2 for the junction layout) that ensures that agents move straight down (in case the input already contains the bit of the operand), or a *reset-TRUE junction* (see Table S2 for the junction layout) that forces agents to take the diagonal path and move to the right. All other junctions are *pass junctions* designed to ensure that agents stay on their respective paths (either straight down or diagonal). Placement of the *reset-TRUE junctions* is determined by checking if the binary number corresponding to the respective column and the operand have common '1' bits. In that case, the *reset-TRUE junction* is placed further down to move the agents fewer steps to the side and arrive at the correct result. For example, column $0000_2$ has no common bits with the operand $0101_2$ and thus the *reset-TRUE junction* is placed immediately at the entrance of the OR network block, guiding the agent to the correct exit at column $0101_2$ ($0000_2 \lor 0101_2 = 0101_2$; purple path 1 in Figure 1b). In contrast, column $0110_2$ has the second bit in common with the operand $0101_2$. Therefore, the *reset-TRUE junction* is placed $0100_2$ rows further down, guiding the agent to the correct exit at column $0111_2$ ($0110_2 \lor 0101_2 = 0111_2$ red path 2 in Figure 1b).

## Step 4: Network representation of 3-SAT

In the binary representation of 3-SAT, each variable is represented by two binary numbers (described in Step 1 above). In order to enable the agents to randomly explore all possible variable assignments, each agent chooses randomly between either of these two numbers and performs the bitwise OR operation for its current value with that number as an operand. This is achieved by encoding both numbers as operands into the same OR operation block in the network and then replacing the first *reset-TRUE junction* at each input with a junction that is designed such that input agents have a 50% chance to choose each variable assignment by moving either diagonally or straight down. If the first *reset-TRUE junction* is at the first row of the OR operation block, it is replaced by a *split junction*, that is designed to distribute agents arriving from the top or from the left equally between both exits of the junction (see Table S2 for the junction layout). If the first *reset-TRUE junction* is at a lower row of the OR operation block, it is replaced by a *split-top junction*, that is designed to distribute agents arriving from the top equally between both junction exits but keep agents arriving from the left on the diagonal path (see Table S2 for the junction layout).

As an example, we have shown the network encoding of instance $\Phi_1$ in Figure 2a. To illustrate the solving process, we have marked all the possible legal paths in the network for this instance (Figure 2a, green dashed lines). In this network, the variable $x_2$ is represented by the first OR block with the only input $0000_2$, and the operands $1101_2$ and $0010_2$ (corresponding to the assignments $x_2$ = TRUE and $x_2$ = FALSE, respectively). In this first OR block, agents have a 50% chance to choose either operand and arrive at output $1101_2$, or output $0010_2$ (green dashed lines in Figure 2a). That way, agents entering the network stochastically pick either the assignment $x_2$ = TRUE (corresponding to the binary number $1101_2$, encoding fulfilment of clauses $(x_1 \lor x_2)$, $(\neg x_1 \lor x_2)$, and $(\neg x_1 \lor x_2)$) or the assignment $x_2$ = FALSE (corresponding to the binary number $0010_2$, encoding fulfilment of clause $(x_1 \lor \neg x_2)$). The output of the first OR operation block in row 13 of the network acts as the input for the next 'OR' operation block that represents variable $x_1$. There, agents randomly pick the binary number operands $1010_2$ ($x_1$ = TRUE) or $0101_2$ ($x_1$ = FALSE). Thus, agents that arrived at column $0010_2$ will exit the second OR operation block at columns $0111_2$ or $1010_2$, while agents that arrived at column $1101_2$ will exit at columns $1101_2$ or $1111_2$.

By observing column $1111_2$, the satisfiability of the instance can be checked. If agents can find a way through the network and exit at column $1111_2$, this indicates that there exists one combination for assignments to $x_1$ and $x_2$ that satisfies the formula. During the operation of a network device, the solution can be read out by following the legal path leading to column $1111_2$. In the example given in Figure 2a, the solution is $x_1$ = TRUE ($1010_2$) and $x_2$ = TRUE ($1101_2$).

## Step 5: Network optimization

Finally, we optimized the network by reducing its physical size before nanofabrication. A smaller network is cheaper to fabricate and can be explored by fewer agents in a shorter time. Specifically, we first replaced the single entrance at the top by two separate entrances, placed at the two columns that represent the two binary numbers for the first variable (see Figure 2a (entire network) compared to Figure 2b (optimized version of network in 2a). By doing this, we could remove the variable space for the first variable, reducing the travelling distance for all agents. Moreover, with two entrances, more agents can enter the network at the same time, increasing operation speed. Second, we truncated the rows to the left of the first variable, because that part of the network is not explored by agents moving correctly according to the network design. Third, recognizing that we need to record the network output only at the one column that represents the all-one binary number (target exit, see step 4), we removed all columns to the right of that column. The optimized network encoding instance is significantly smaller than the original design (compare Figure 2 a and b) but still explores all non-trivial, possible solutions, and its creation requires no prior knowledge of the solution. The network layout for all the instances can be found in the Supplementary (Figures S1-S4).

**Experimental demonstration**

In order to demonstrate the function and to analyse the error rates of the 3-SAT networks we used actin filaments as biological agents that were propelled by myosin II motor fragments (heavy meromyosin) along the surface in gliding motility assays [19-23]. Actin filaments are highly flexible with a very low escape rate from nanochannels [21] when propelled by fast myosin II molecular motors from skeletal muscle. With a speed of up to 15 µm/s, a high value among biomolecular motors [24], fast computation is also enabled using this motor-filament combination [20,21,24-30]. Reliable guiding of agents through the network was enabled by two features: i) physical guiding, provided by channels that are patterned into a layer of polymer to form a network; ii) chemical guiding, achieved by selectively functionalizing only the floor of the channels to enable actin-myosin motility. As a result of (i) and (ii), the filaments can only attach to the channel floors where they are propelled by motors and move, guided by the channel walls [21]. Recent improvements of the actomyosin function and longevity in nanofabricated networks [31] have been instrumental in facilitating effective use of this motor-filament system for the present purpose.

The general layout of the networks implementing the algorithm explained above is shown in Figure 3a with details of the key junctional elements given in Figure 3b-f (see also Table S2). The width of the network channels was 100 nm with a wall height of 400 nm. Two loading zones (rows of large heart shapes in Figure 3a) were placed to guide filaments to enter the network from columns representing numbers corresponding to one variable (here at column 2 and column 13, see Figure 3a and g). One row of rectifier structures was introduced at the bottom of the network to stop filaments from re-entering the network. One long outer channel was introduced between exits and loading zones to recirculate the filaments.

The devices were then operated employing an in vitro motility assay [19], where fluorescently labelled actin-filaments from the solution were collected and landed in the loading zones from where they were guided by non-labelled myosin-II motor fragments to explore the network. For reading out the results, fluorescence time-lapse movies of the process were recorded (see details in **Materials and Methods**). The data in Figure 4a, based on such recordings, show the standard deviation projection of the resulting time series. In this projection, more frequently travelled paths appear brighter. Each straight part of the paths at the bottom represents one exit. Quantitatively, the distribution of the filaments leaving each exit was obtained by counting the filaments in image sequences where each filament was tracked (blue bars, Figure 4b). Similarly, the error rate at *pass junctions* was determined to be 2.20%±0.09% by counting the fraction of agents that take a turn at pass junctions (567) divided by the total number of pass junction crossing events (25748).

In our algorithm, the satisfiability of the formula is determined by whether there are filaments that leave the network at the target exit, i.e. at column 15 corresponding to $1111_2$ for problems $\Phi_1$ (Figure 2a) and $\Phi_2$, or column 31 corresponding to $11111_2$ for problem $\Phi_3$ and $\Phi_4$, respectively. However, in a real device, filaments may inadvertently turn incorrectly at *pass junctions* and exit at incorrect columns (not following the rules of the networks) thus creating a background of errors in the filament distribution histogram. Therefore, we need to be able to distinguish correct results for a given exit, where filaments arrive while following the rules of the junctions in the network, and incorrect results, where filaments arrive at a given exit due to wrong turns.

**Quantifying the significance of computation using statistical analysis**

NBC is inherently stochastic and can deliver answers only with a confidence level lower than 1 (100 %). Calculating the confidence level is thus an integral part of solving a problem instance. To quantitatively assess the computation result from NBC in view of its stochastic nature, we developed a statistical method to estimate the probability that the results measured for each network indicate that the corresponding problem instance is satisfiable. For the networks encoding 3-SAT instances, this is equivalent to estimating the probability that the target exit corresponds to a correct result.

The statistical analysis is described in detail in the supplementary material. Briefly, we begin by estimating the fraction of useful agents (those that do not make wrong turns at any *pass junction*) from the average number of *pass junctions* that agents will travel through and the fractional error per *pass junctions* (*f*, see the Supplementary Materials and Methods). We then used the expected number of agents that follow a legal path and an illegal path under these conditions. Then, using the normal approximation of a binomial distribution we arrive at a z-score based calculation of a significance level that also corresponds to a certain confidence level (see the Supplementary Materials and Methods and Figure S5 for further details). Comparing the experimental readout to the expected values (Figure 4b) resulting from our analysis, instances $\Phi_1$ and $\Phi_3$ are satisfied with a 99 % confidence level. Conversely, our results suggest that the instances $\Phi_2$ and $\Phi_4$ are unsatisfied with a 99 % confidence level.

**Discussion**

We here demonstrate the successful application of NBC to solve four instances of the key NP-complete problem, 3-SAT. To achieve this, a highly multidisciplinary approach was necessary. First, we developed an algorithm that encodes 3-SAT into network format. We used this algorithm to encode four 3-SAT instances into four networks of channels. After nanofabrication and fine-tuned selective functionalization of the network channels, myosin driven actin filaments effectively searched the entire network. Finally, we devised a statistical approach to determine whether a significant number of filaments ($p < 0.01$) arrived at the target exit, indicating whether the respective 3-SAT instance was satisfiable or not. All networks were solved correctly with 99 % confidence, demonstrating that the network algorithm works correctly. This is a key step towards broad general applicability of NBC because NP-completeness is usually proven mathematically by finding a polynomial conversion to SAT [8]. Furthermore, polynomial conversions to 3-SAT exist for a wide set of practically important NP-complete problems [9].

To estimate requirements for scaling our 3-SAT NBC algorithm to solving larger 3-SAT instances, we consider a benchmark instance with 8 clauses and 10 variables, which results in a network similar in size to a current electronic CPU: The network will have ~ 2300 rows corresponding to a chip of ~24×24 mm², which is certainly feasible to fabricate and poses no problem with respect to the possibility of exploration by the actomyosin system in view of the recently achieved longevity of motile function for several hours to almost a day [31] and a gliding velocity of ~ 10 µm/s. To avoid that a majority of filaments crossing the network makes errors, the error rate at pass junctions needs to be below 0.07% [15], a factor of 4 better than the best error rate reported for a NBC calculation so far [15] and certainly achievable with error-free 3D junctions[32]. To ensure a sufficient number of agents passed this network, ~50000 actin filaments are required[14,33], which corresponds to ~$5*10^{-10}$ mg actin, available at very low cost. Third, agents need to be detected efficiently at exits such as electrical/optical sensors [34-36]. It is important to note that energy consumption would not pose any limitation due to the energy efficiency of the biomolecular motors [14]. Concluding from these considerations, the major limiting factor for further scale-up of the SAT network encoding presented here, is the unary encoding. The unary encoding requires that the networks grow exponentially with the number of clauses in the SAT problem. For example, each OR operation block needs 8 rows for 3 clauses, 32 rows for 5 clauses and 256 rows for 8 clauses. The number of OR operation blocks required is one less than the number of variables. Thus, the network grows exponentially with the number of clauses and linearly with the number of variables. The reason behind the exponential growth of the network size is that information is stored as the position of the agents within the network. Fundamentally, this could be avoided, if the information about the agent path could be stored on the agents themselves (tagging). A reasonable assumption is that only one *split junction* is required per variable for a tagging-based device. This would significantly reduce the size of the chip and vice versa, a much larger instance can be encoded in a small chip, e.g., a chip of size 24×24 mm$^2$ would accommodate a 3-SAT instance with ~5 million (2300$^2$) variables, with the number of clauses only being limited by the number of available tags.

## Materials and Methods

**Fabrication of computational networks** The fabrication of the computational networks was done on a single side polished 150 mm Si (100) wafer. A 70 nm thick SiO$_2$ on the Si-substrate was made by dry thermal oxidation. The layer growing was carried out under oxygen atmosphere with 3% HCl. After an O$_2$ plasma cleaning and surface activation steps for 30 seconds, a resist layer (ALLRESIST AR-P 6200:10) was spin-coated onto the SiO$_2$, to a thickness around 400 nm and hard-baked at 180°C for 2 minutes. The network was patterned by electron beam lithography (Vistec SB254) with a dose of around 100 µC/cm$^2$. The AR-P 6200:10 was developed in 600594 developer for 60 seconds, rinsed with IPA and flushed in a conductance-controlled DI-water bath. Finally, the wafer was spun dry and separated into 1 cm$^2$ chips.

To make the surface suitable for protein anchoring and support of motility, silanization with trimethylchlorosilane (TMCS; Sigma-Aldrich Sweden AB, Stockholm, Sweden) was performed. In detail, the samples were O$_2$ plasma etched (Plasma Preen II-862, Plasmatic Systems, Inc., North Brunswick, NJ) at 5 mbar for 45s to remove resist residues and remove the adsorbed water molecules. The etching was performed within a Faraday cage to achieve isotropic etching so the sidewalls in the nanostructures were etched as well [31]. The samples were then immersed in DI-water for 5 minutes to generate as many as possible hydroxyl groups on the surface for later silanization. The silanization was then performed in a controlled chemical-vapour deposition (CVD) system following the method in the reference [37].

**Protein preparations** Rabbits were sacrificed in accordance with the ethical guidelines and procedures approved by the Regional Ethical Committee for Animal Experiments in Linköping, Sweden (ref.no.: 73-14). Myosin-II was isolated from rabbit hind leg fast muscles [38,39], followed by digestion with Tosyl-L-lysine-chloromethyl ketone hydrochloride (TLCK) treated α-chymotrypsin to obtain heavy meromyosin (HMM) [40]. Actin was isolated from rabbit back muscles [41]. Fluorescence labelling of actin filaments was achieved using Rhodamine-phalloidin [42].

**Solutions and chemicals** For all solutions used in the in vitro motility assay type of experiments, low-ionic strength solution (LISS) was first prepared as a basal component, containing 1 mM magnesium chloride ($MgCl_2$), 10 mM 3-(N-morpholino) propanesulfonic acid (MOPS), 0.1mM $K_2$-ethylene glycol tetra-acetic acid (EGTA), with a pH of 7.4 and ionic strength of 15 mM. Wash solution was prepared by adding 1mM Dithiothreitol (DTT) and 50mM potassium chloride (KCl) to the LISS solution (final concentrations). Finally, assay solution was prepared by adding 0.2 mg/ml creatine phosphokinase (CPK), 2.5 mM creatine Phosphate (CP), 1 mM magnesium adenosine triphosphate (MgATP), an oxygen scavenger mixture (GOC): 3mg/ml Glucose : 0.1 mg/ml Glucose Oxidase, 0.02 mg/ml Catalase, 10 mM DTT and 45 mM (KCl) to the LISS solution, giving a final ionic strength of 60 mM [43,44].

**In vitro motility assays** HMM, fluorescently labelled actin and non-fluorescent blocking-actin [45] were diluted in the wash solution. To perform the in vitro motility assay, flow cells were built using a glass coverslip and a nanostructured chip on top, separated with Parafilm spacers [21,37]. Flow cells were incubated using the following steps: 120 µg/ml HMM (5 min); 1mg/ml bovine serum albumin (2 min); wash solution (washing, 1x); 0.5 µM blocking-actin (2 min); 1mM MgATP dissolved in wash buffer (2 min); wash solution (washing, 1x); 15 nM rhodamine-phalloidin labelled actin filaments (2 min); wash (1x) and final incubation with assay solution (2 min). The IVMAs for all nanostructured chips were performed within the temperature range 24–26 ºC using a temperature-controlled objective coil. Usually, the flow cells were sealed with silicone vacuum grease during experimental observations.

**Imaging methods** IVMA experiments were imaged using an inverted fluorescence microscope (Zeiss Axio Observer.D1), mounted with either 63x or 40x objective (numerical aperture: 1.4 or 1.3) and Hamamatsu EMCCD camera (512 x 512 pixels). Image sequences were recorded at an exposure time of either 200 ms (4.96 frames/second) or 400 ms (2.5 frames/second). A mercury short-arc lamp (OSRAM) was used along with a filter (cyanine-3) for the excitation of Rhodamine fluorescence. Images were then analysed with Fiji software [46].

46    Schindelin, J. *et al.* Fiji: an open-source platform for biological-image analysis. *Nat Methods* **9**, 676-682, doi:10.1038/nmeth.2019 (2012).

**Acknowledgments**

**Funding:** This work was funded by:
European Union's Horizon 2020 research and innovation programme under grant agreement No 732482 (Bio4Comp)
NanoLund, Lund University SE

**Author contributions:** T.K., H.L., S.D., A.M., H.K. and D.R., conceived the study. J.Z., T.K. and H.K. developed the algorithm. C.M., J.Z. and P.S. nanofabricated the computation networks. J.Z., P.S., R.L. and A.S. performed actin-myosin in vitro motility assays and analyzed the respective data. J.Z. and T.K. performed the statistical analysis. J.Z., A.S. and T.K. wrote the manuscript with contributions from all authors. All authors approved the final version of this manuscript.

**Competing interests:** Authors declare that they have no competing interests.

**Data and materials availability:** All data needed to evaluate the conclusions in the paper are present in the paper and/or the Supplementary Materials. Additional data related to this paper may be requested from the authors.

**Figures and Tables**

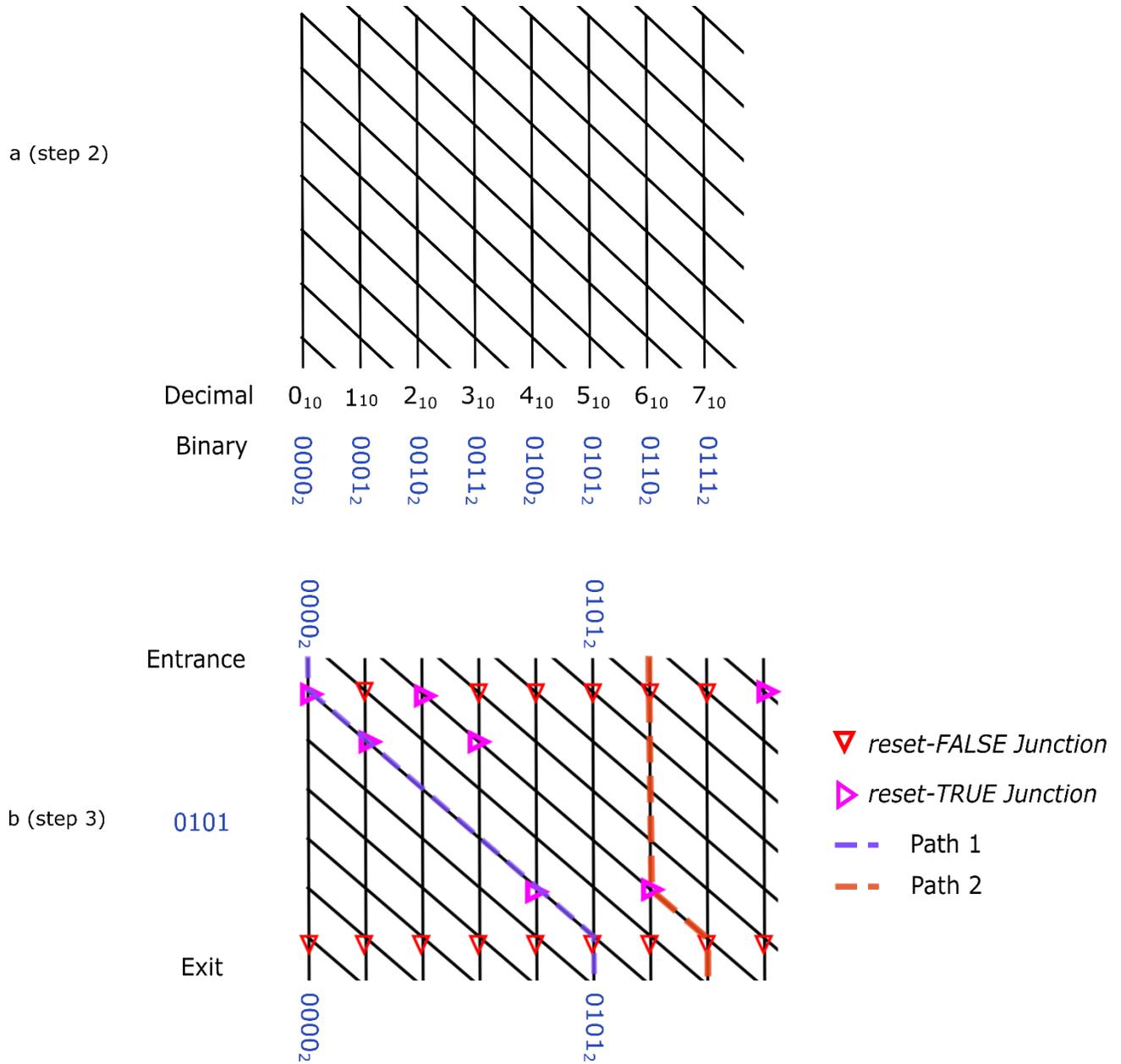

Figure 1 Network encoding examples. **a** Example network shows numbers represented by columns ranging from $0000_2$ to $0111_2$. **b** Example network encoding bitwise 'OR' with the operand $0101_2$ for all inputs between $0000_2$ and $1000_2$. Path 1 shows the agent entering the network at column $0000_2$ where a *reset-TRUE junction* is introduced; no shift is required. Path 2 shows the agent entering the network at column $0110_2$ where there is a common '1' bit with operand $0101_2$ at the 2nd most significant digit. The *reset-TRUE junction* is shifted by four ($2^2$) rows down.

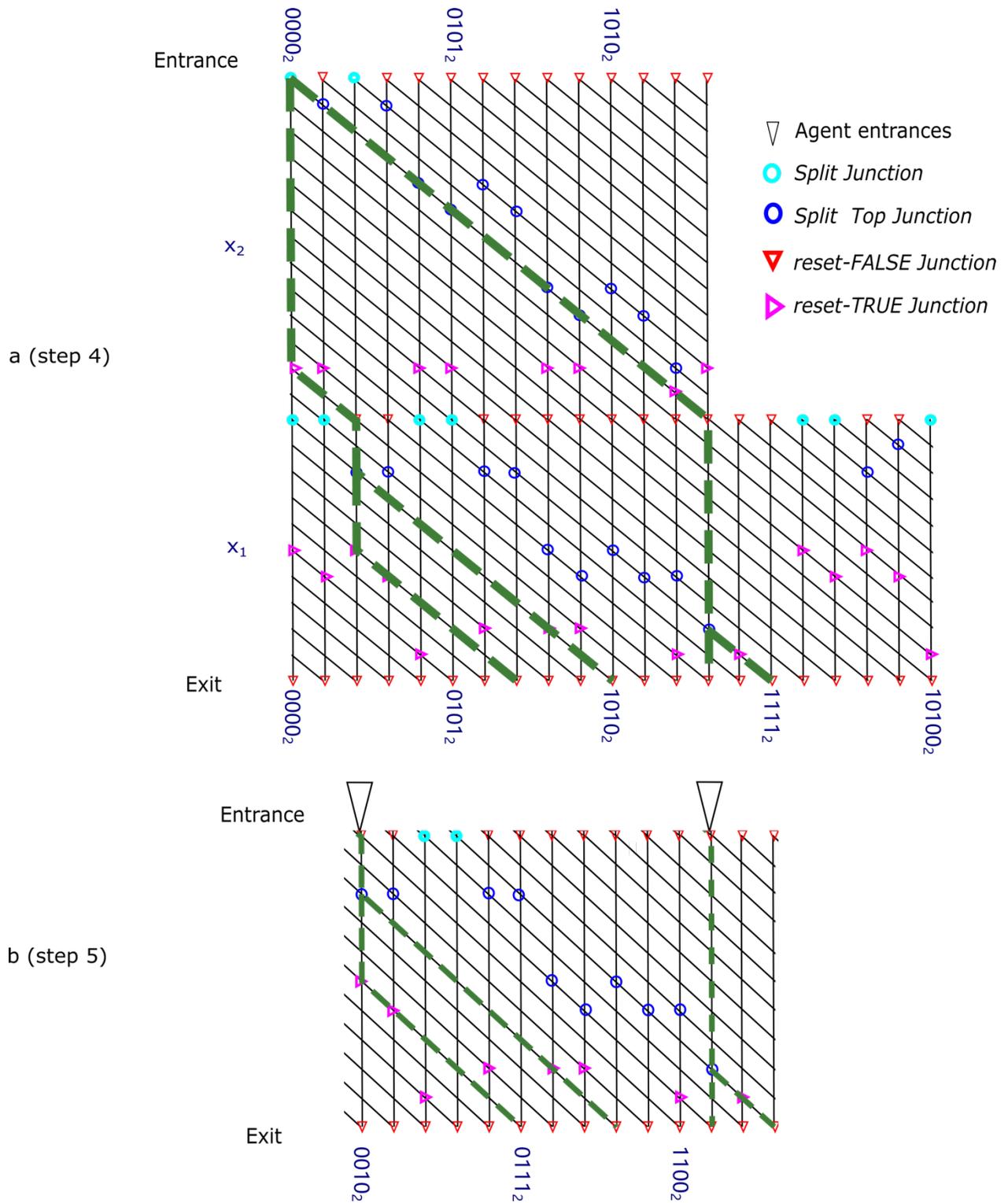

**Figure 2 Network encoding for instance $\Phi_1$. a** Full layout design of network encoding for instance $\Phi_1$. The green paths are all the possible paths in the network where agents can travel according to the network design. The corresponding detailed physical layout of the junctions is shown in Figure 3. **b-f**, The optimized spatially encoded network design for 3-SAT instance $\Phi_1$ following Step 5.

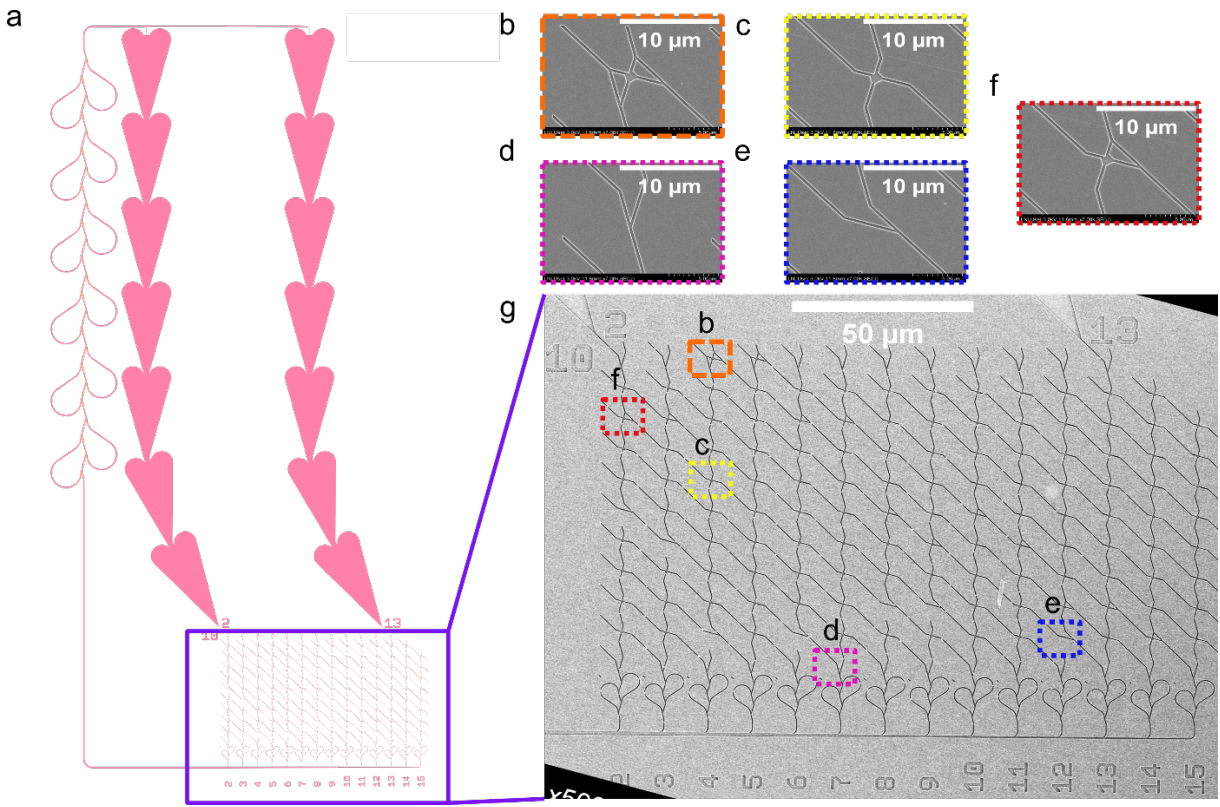

**Figure 3 Physical network layout for solving instance $\Phi_1$.** **a** Device layout. Schematic of the actual device used for experiments. Filaments enter the network via the loading zones at two entrances on the top. Filaments that exit the network at the bottom are collected via the recirculation path and guided to loading zones for multi-time operations; **b** SEM image of a *split junction*. **c**, SEM image of a *pass junction*. **d** SEM image of a *reset-FALSE junction*. **e** SEM image of a *reset-TRUE junction*. **f** SEM image of a *split top junction*. **g** SEM image of the overall network.

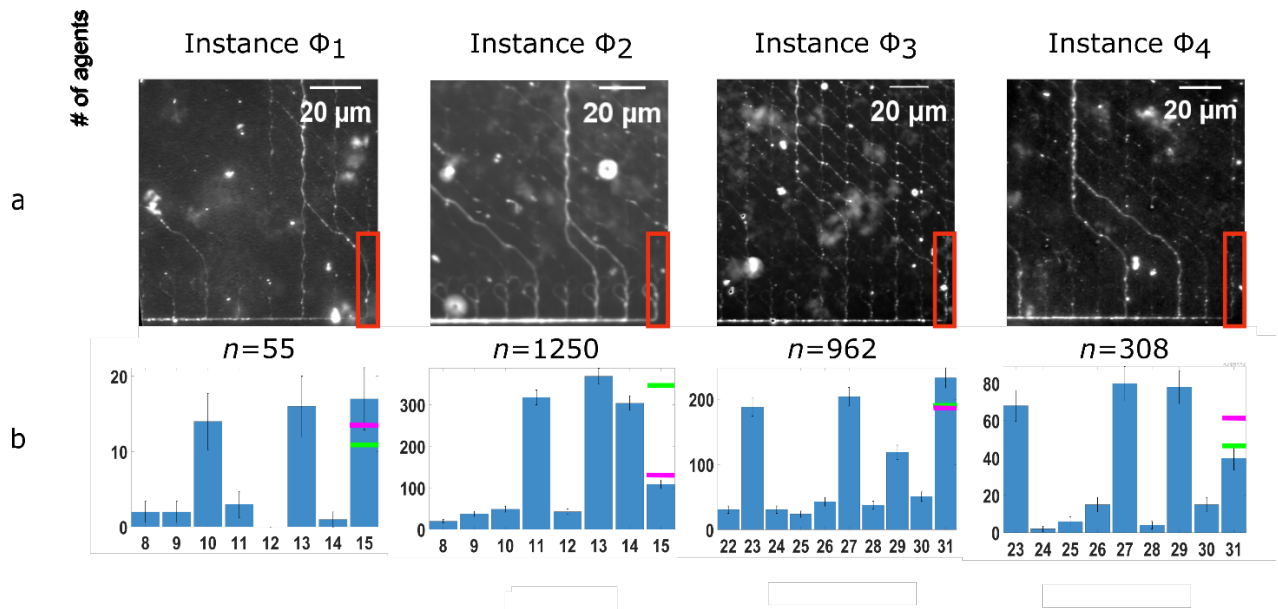

**Figure 4 Experimental read-out for four instances of 3-SAT ($\Phi_1, \Phi_2, \Phi_3, \Phi_4,$).** a Standard deviation of projections of one thousand typical fluorescence micrographs (duration of 400s) of actin filaments moving through the device. b Experimental results obtained from manually counted actin filaments passing the exits at the bottom. Total counted events ($n$) are displayed. Error bars represent the counting error ($\sqrt{n}$). For the target column (15 for $\Phi_1, \Phi_2$, and 31 for $\Phi_3, \Phi_4$, shown in red boxes in a), values above the magenta bar are significantly correct exits ($p < 0.01$, corresponding to 99 % confidence level, see the Supplementary Materials-1), values below the green bar are significantly incorrect exits ($p < 0.01$, see the Supplementary Materials-1). Both criteria are mutually exclusive, otherwise, the result is not conclusive. For example, $\Phi_2$ would be inconclusive if column 15 had received more than 132 and less than 333 filaments.

**Table 1 Binary representations of instance $\Phi_1$.** Binary numbers forming by the logical values from all four clauses. Each variable ($x_1$ and $x_2$) will generate one pair of binary numbers when assigned TRUE or FALSE.

|  | Assignment | $(x_1 \vee x_2)$ | $(\neg x_1 \vee x_2)$ | $(x_1 \vee \neg x_2)$ | $(\neg x_1 \vee x_2)$ | Binary |
|---|---|---|---|---|---|---|
| $x_1$ | TRUE | fulfilled | Not fulfilled | fulfilled | Not fulfilled | $1010_2$ |
|  | FALSE | Not fulfilled | fulfilled | Not fulfilled | fulfilled | $0101_2$ |
| $x_2$ | TRUE | fulfilled | fulfilled | Not fulfilled | fulfilled | $1101_2$ |
|  | FALSE | Not fulfilled | Not fulfilled | fulfilled | Not fulfilled | $0010_2$ |

**Table 2 Truth table for the instance formula $\Phi_1$.** The binary representation of both variables ($x_1$ and $x_2$) and bitwise 'OR' operation result of all the possible combinations of variable assignments. The assignments $x_1$ = TRUE and $x_2$ = TRUE represent the solution for $\Phi_1$

| $x_1$ | $x_2$ | Binary $x_1$ | Binary $x_2$ | Bitwise 'OR' of Binary $x_1$ and Binary $x_2$ | Output |
|---|---|---|---|---|---|
| TRUE | FALSE | $1010_2$ | $0010_2$ | $1110_2$ | FALSE |
| FALSE | TRUE | $0101_2$ | $1101_2$ | $1101_2$ | FALSE |
| TRUE | TRUE | $1010_2$ | $1101_2$ | $1111_2$ | TRUE |
| FALSE | FALSE | $0101_2$ | $0010_2$ | $0111_2$ | FALSE |

# Supplementary Materials for

## Solving the 3-SAT problem using network-based biocomputation

Jingyuan Zhu[1, 2], †, Aseem Salhotra[3, †], Christoph Robert Meinecke[4], Pradheebha Surendiran[1, 2], Roman Lyttleton[1, 2], Danny Reuter[4, 5], Hillel Kugler[6], Stefan Diez[7], Alf Månsson[1, 3*], Heiner Linke[1, 2, *], Till Korten[7, *]

*Corresponding authors: alf.mansson@lnu.se, heiner.linke@ftf.lth.se, till.korten@tu-dresden.de

**This PDF file includes:**

## Statistical evaluation of the experimental results

To quantify the uncertainty of the computation done by NBC, statistical analysis was performed as described below.

### Estimated number of useful agents (not making any wrong turns)

Agents that are classified as useful (from now on termed "useful agents") should obey the designed traffic rules at all junctions in the network, meaning that they follow a "legal path". However, in practice, some agents do turn incorrectly at *pass junctions*, which results in "illegal paths". Agents following illegal paths (from now on termed "useless agents") can arrive at "incorrect exits". First, we need to estimate the fraction, f, of useful agents:

$$f = (1 - E)^x \tag{1}$$

where $E$ is the fractional error per *pass junction* and $x$ is the number of *pass junctions* passed by the agent. We assume the same error fraction $E$ for all *pass junctions* in the network. See Table S1 for the error fractions measured experimentally for each network.

Thus, the number of useful agents is:

$$N_{useful} = N_{total} * f \tag{2}$$

where $N_{total}$ is the total number of agents which explored the network. Following the same assumptions, the number of useless agents is:

$$N_{useless} = N_{total} * (1 - f) \tag{3}$$

### Expectation for the number of agents taking a specific legal path through the network

Assuming *split junction*s with a split ratio of 50%, all useful agents are equally likely to visit all legal paths. For a network encoding a SAT instance with $m$ variables and $c$ clauses, there are $2^m$ possible legal paths, and $2^c$ total exits. The minimum fraction of agents per legal path is then approximately:

$$P_{correct} = \frac{f}{2^m} \tag{4}$$

where $f$ is the fraction of useful agents among all the agents. This is the minimum number of useful agents we expect to arrive at a correct exit. Note that a larger number of useful agents can arrive at a correct exit, if more than one legal path leads to said exit.

### The expectation for errors at the target exit

Two principal assumptions are central to allow error estimation at the target exit:

1. Useless agents which take wrong turns more than once will not be considered (no secondary errors). This assumption can be justified, because the current fractional error per *pass junction* (0.0220±0.0009), results in a negligible secondary fractional error (0.0005±0.00127) that will not significantly contribute to the final agent distribution.

2. Useless agents, taking a wrong turn within each variable's OR block are assumed to be evenly distributed among all columns except the target column.

Based on the above assumptions, useless agents arriving at the target exit will have two contributions: 1. The useless agents from the same column of the exit from the previous variable's OR block (green arrow in Figure S1) and 2. Half of the useless agents arriving at *split junctions* or *split top junctions* on the diagonal line in each variable's OR block, which leads to the target exit (red arrow in Figure S1).

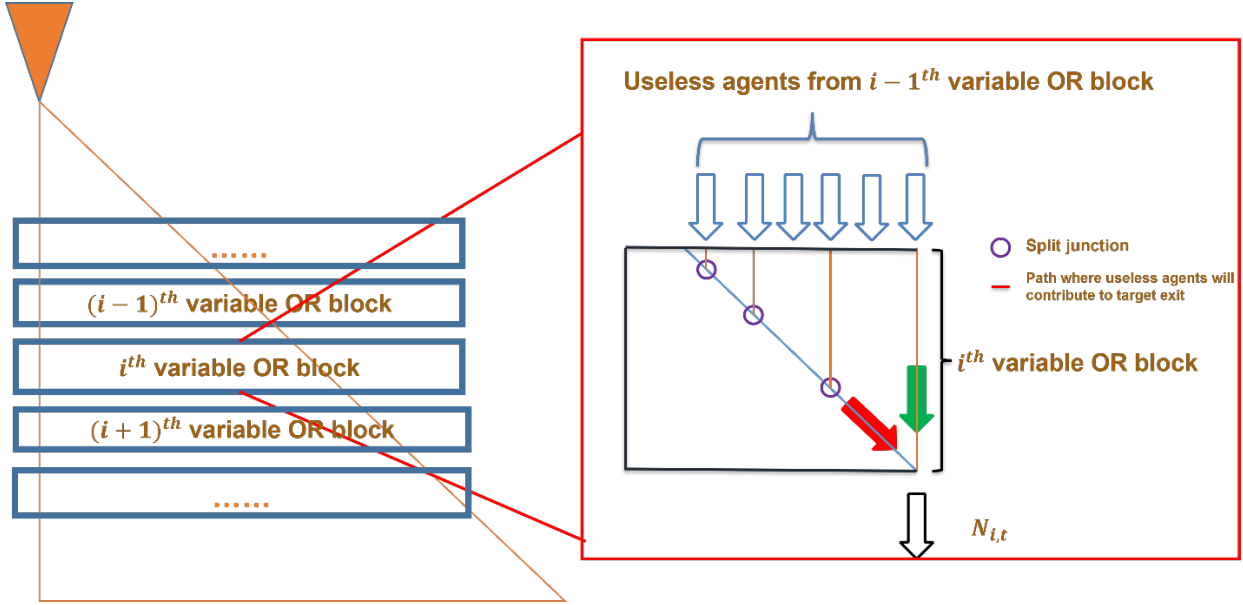

**Figure S1** Schematic of the iteration of useless agents at target exit for the $i^{th}$ variable's OR block.

Based on these considerations, we estimate the number of useless agents arriving at the target column in four steps:

i) The number of useless agents, $N_{i,t}$, at the target column, in the $i^{th}$ variable's OR block will be:

$$N_{i,t} = N_{i-1,t} + N_{i,d} * \frac{1}{2} \qquad (5)$$

where $N_{i,d}$ is the number of useless agents reaching the *split/split top junction*s on the diagonal line which leads to the target column within the $i^{th}$ variable's OR block (red arrow in Figure S1) and $N_{i,t}$ is a recursive sequence.

ii) Now, in order to estimate $N_{i,d}$, we need to find the number of *split/split top junction*s in the diagonal line of interest that leads to the target column in the i$^{th}$ variable's OR block. The target column has all ones in the binary column number. As described in the main text, this means that the encoding algorithm places a *split/split top junction* on the diagonal line of interest in each column which has a 0 where there is a '1' in one of the two binary operands of the respective OR block and '1' everywhere else. For example if we look at Figure S2 that represents the OR block for variable $x_1$, which encodes the operands $1010_2$ and $0101_2$, we see *split/split top junctions* in columns 7 ($0111_2$), 13 ($1101_2$), and 5 ($0101_2$), corresponding to 0 in the position of the 1s in the first operand as well as 11 ($1011_2$), 14 ($1110_2$), and 10 ($1010_2$), corresponding to 0 in the position of the 1s in the second operand.

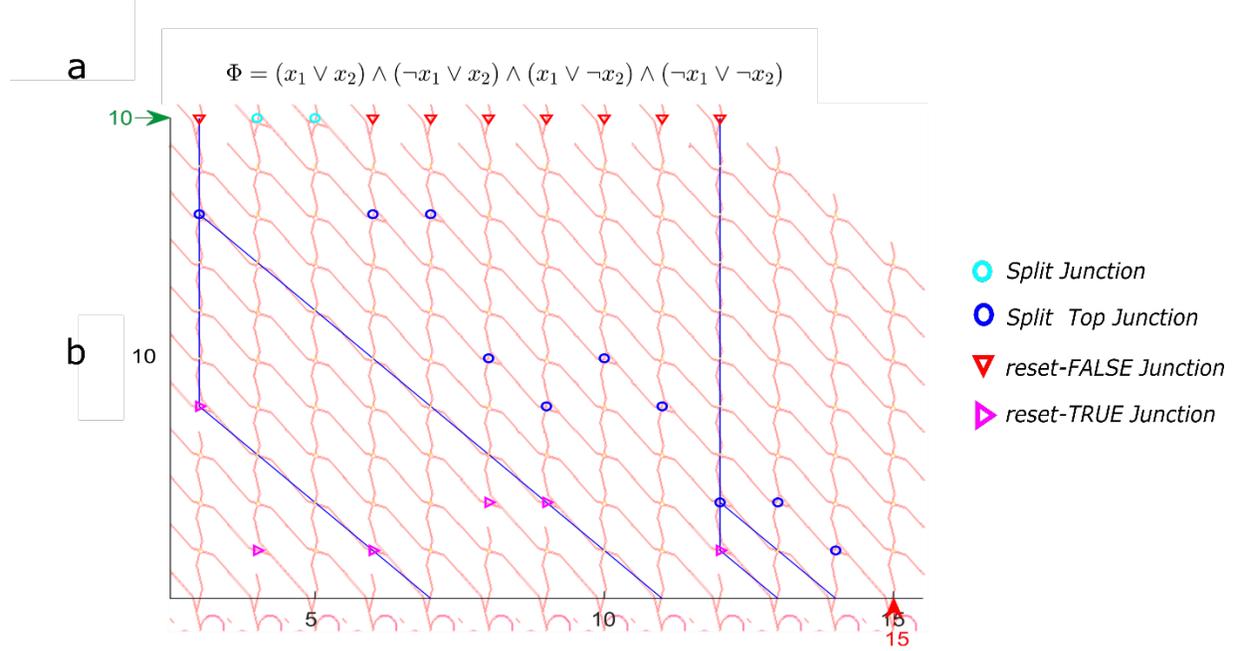

**Figure S2** (**a**), the Boolean formula of instance $\Phi_1$. (**b**), physical network layout for instance $\Phi_1$.

Thus, the number of *split junction*s on the diagonal leading towards the target column is calculated as follows:

$$N_{sj} = 2^{a_i} + 2^{b_i} - 2 \qquad (6)$$

where, $a_i$ and $b_i$ is the number of '1's in the corresponding binary operands encoded in the $i^{th}$ variable's OR block.

According to assumption 2 above, the useless agents distribute evenly in the rest of the network. Therefore, the number of useless agents reaching split junctions on the diagonal is:

$$N_{i,d} = \frac{N_{sj}}{C} * N_{i-1,nt} \qquad (7)$$

where C is the number of columns in the network and $N_{i-1,nt}$ is the number of useless agents in the $(i\text{-}1)^{th}$ variable's OR block, except for the target column (the subscript nt: "not target"):

$$N_{i,nt} = N_i - N_{i,t} \qquad (8)$$

where $N_i$ is the number of useless agents in the $i^{th}$ variable's OR block.

iii) By combining equations 5 - 8, the number of useless agents at the target column would be:

$$N_{i,d} = \frac{2^{a_i} + 2^{b_i} - 2}{C} * (N_{i-1} - N_{i-1,t}) \tag{9}$$

$$N_{i,t} = N_{i-1,t} + \frac{2^{a_i} + 2^{b_i} - 2}{C} * (N_{i-1} - N_{i-1,t}) * \frac{1}{2} \tag{10}$$

By dividing equation (10) with $N_{useless}$ (number of useless agents in the whole network), we get the fraction of all useless agents arriving at the target column of the $i^{th}$ variable's OR block:

$$R_{i,t} = R_{i-1,t} + \frac{2^{a_i} + 2^{b_i} - 2}{C} * (1 - R_{i-1,t}) * \frac{1}{2} \tag{11}$$

iv) Based on Eqs. 3, 4 and 11, we estimate thresholds for the target column used to decide if the SAT instance encoded in the network has a solution or not:

$$T_s = P_{correct} + P_{incorrect} = \frac{f}{2^m} + (1-f) * R_{i,t} \tag{12}$$

$$T_{ns} = P_{incorrect} = \frac{N_{i,t}}{N_{total}} = \frac{N_{error} * R_{i,t}}{N_{total}} = (1-f) * R_{i,t} \tag{13}$$

where the threshold $T_s$ represents the scenario when the target column is a correct exit (the SAT instance has a solution), the threshold $T_{ns}$ represents the scenario when the target column is not a correct exit (the SAT instance has no solution).

**Hypothesis test**

Based on the thresholds $T_s$ and $T_{ns}$ calculated as described above, hypothesis tests are performed to evaluate if the experimental data are significantly above or below the respective threshold values assumed under a null hypothesis with one-tail alternative hypotheses. The number of agents at one given exit after one agent exploration is a binary random variable $Y \in \{0,1\}$. The probability that $Y=1$ is given by $p \in [0,1]$. For $N$ agent explorations, we then have a mean value, $NE(Y) = Np$ and a variance $\sigma^2 = Np(1-p)$ [1,2]. The normal distribution approximates the binomial distribution, if $N$ is large (when $Np > 5$). In our case, the distribution of the number of agents in the correct exit is therefore approximately normally distributed with mean value $Np$ and variance $Np(1-p)$.

To evaluate if the observed target column belongs to one of the two possible distributions (either there exists a legal path to the target exit or not), we run Z-tests with $P_t$ corresponding to the observed fraction of all filaments that exit at the target exit.

In practice, we will perform two hypothesis tests to determine whether or not the SAT instance encoded by the network has a solution:

1. Does the observed number of filaments at the target exit suggest that a legal path to the target exit exist?

Null Hypothesis 0: $H_0 : P_t = T_{ns}$; alternative hypothesis $H_1: P_t > T_{ns}$, i.e. the column $t$ is one of the correct exits and the SAT instance is satisfiable.
The p-value, that the Null hypothesis is true is derived from

$$Z = \frac{P_t - T_{ns}}{\sqrt{P_t(1-P_t)/N}} \qquad (14)$$

if the resulting p-value is below the chosen significance level α, $H_0$ is rejected and we assume that the SAT instance is satisfiable.

2. Does the observed number of filaments at the target exit suggest that no legal path to the target exit exist?

    Null Hypothesis 0: $H_0 : P_t = T_s$; alternate hypothesis $H_1: P_t < T_s$.
    The p-value is calculated using equation 14 with $T_{ns}$ substituted by $T_s$. If the p-value is below the chosen significance level α, $H_0$ is rejected and we assume that the SAT instance is not satisfiable.

For the observed fraction of agents arriving at the target exit ($P_t$),

    The probability that instance is satisfied: $P(P_t > T_{ns})$

    The probability that instance is not satisfied: $P(P_t < T_s)$

Overall, the instance can only be satisfied or unsatisfied, the result should be the associate result of the two probabilities. By taking α = 0.01, we can ask the significance level to be 99%.

    Satisfiability= XOR ($P(P_t > T_{ns}), P(P_t < T_s)$).

If either none or both of the two hypothesis tests fails, the experiment indicates a non-conclusive result.

The data of four 3-SAT instances from experiments are then analyzed and displayed in Table S1.

**Table S1. Statistical evaluation of experimental results for the four instances of 3-SAT**

| Instance | Fractional error at *pass junctions* E | Total number of pass junction crossing events | Number of agent runs, $N=$ | Observed number of agents in target column | Fraction of agents in target column, $P_t$ | Z-score ($T_{ns}$) | p-value ($T_{ns}$) | Z-score- ($T_s$) | p-value ($T_s$) | Significance level $\alpha$ | Result |
|---|---|---|---|---|---|---|---|---|---|---|---|
| $\Phi_1$ | 0.023±0.0079 | 353 | 73 | 17 | 0.233 | -0.555 | 0.289 | 2.804 | 0.003 | 0.99 | Satisfiable |
| $\Phi_2$ | 0.035±0.0021 | 7500 | 1526 | 108 | 0.071 | -25.6 | 0.000 | -0.353 | 0.638 | 0.99 | Not satisfiable |
| $\Phi_3$ | 0.016±0.0011 | 13949 | 1279 | 233 | 0.181 | 0.020 | 0.508 | 4.94 | 0.000 | 0.99 | Satisfiable |
| $\Phi_4$ | 0.021±0.0023 | 3946 | 410 | 40 | 0.098 | -4.69 | 0.000 | -1.08 | 0.860 | 0.99 | Not satisfiable |

# Figure S3

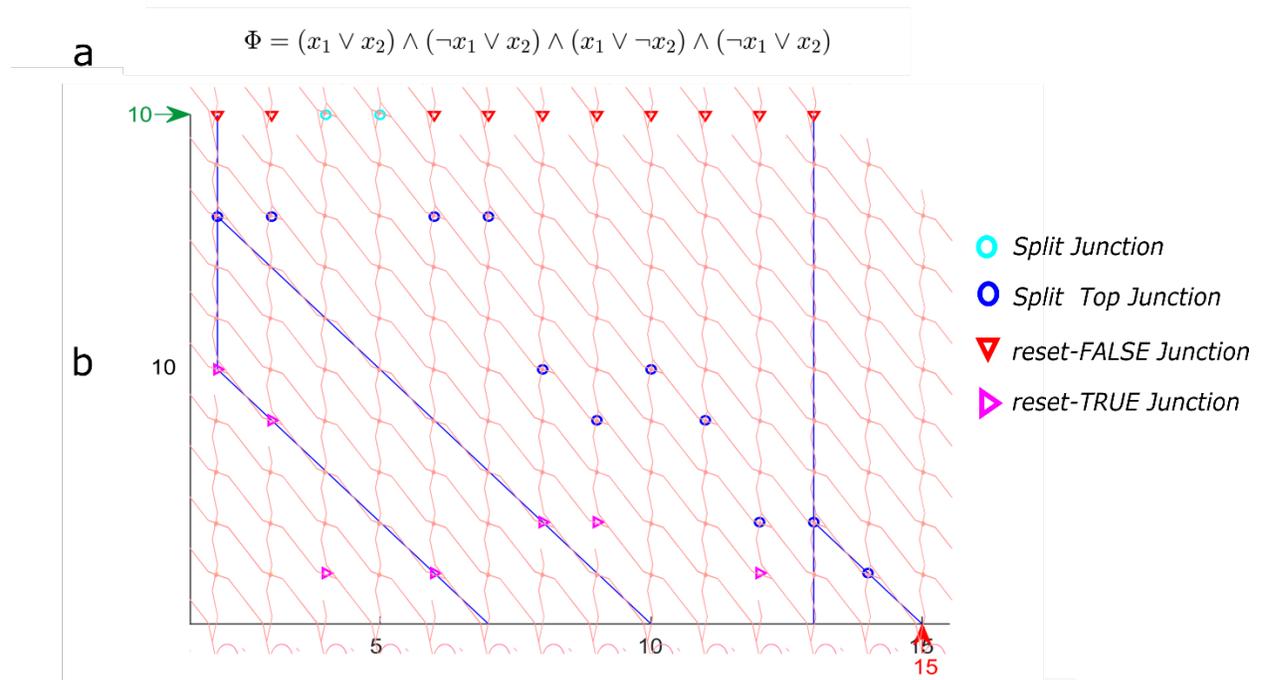

(**a**), the Boolean formula of instance $\Phi_2$. (**b**), physical network layout for instance $\Phi_2$.

# Figure S4

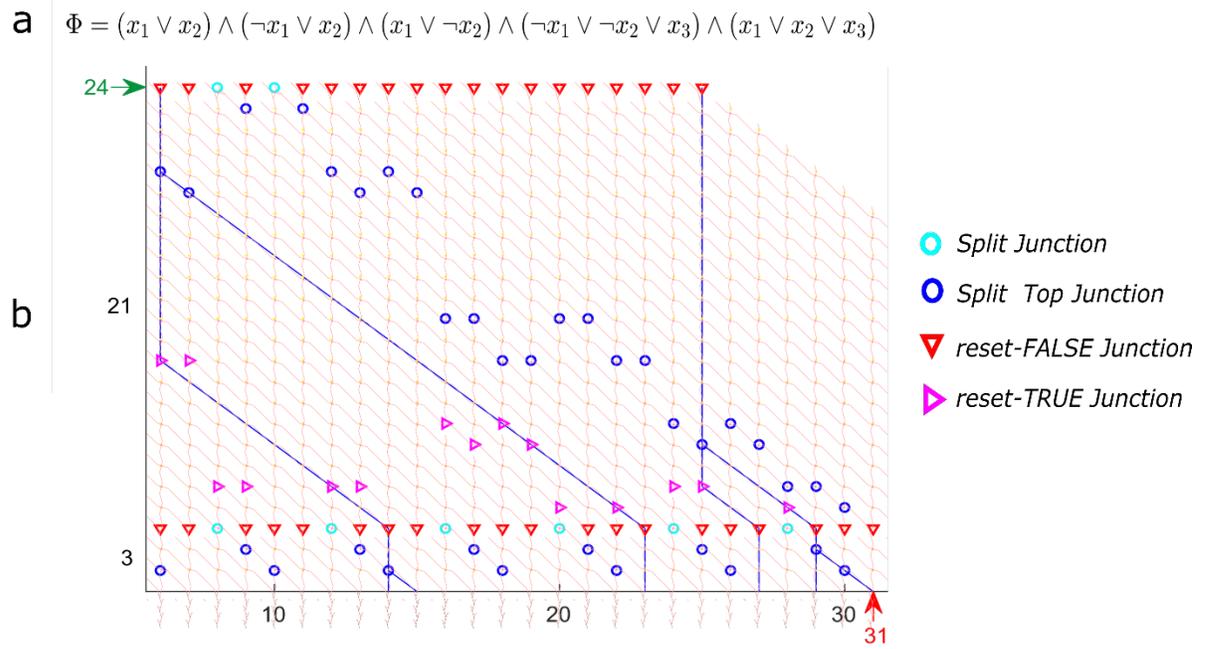

(**a**), the Boolean formula of instance $\Phi_3$. (**b**), physical network layout for instance $\Phi_3$.

# Figure S5

**a** $\Phi = (x_1 \vee x_2) \wedge (\neg x_1 \vee x_2) \wedge (x_1 \vee \neg x_2) \wedge (\neg x_1 \vee \neg x_2) \wedge (x_1 \vee x_2 \vee x_3)$

**b**

- ○ Split Junction
- ● Split Top Junction
- ▽ reset-FALSE Junction
- ▷ reset-TRUE Junction

(**a**), the Boolean formula of instance $\Phi_4$. (**b**), physical network layout for instance $\Phi_4$.

# Table S2. Different junction geometries used in the network encoding of 3-SAT

| Junction geometries and possible exploration paths | Description |
|---|---|
| 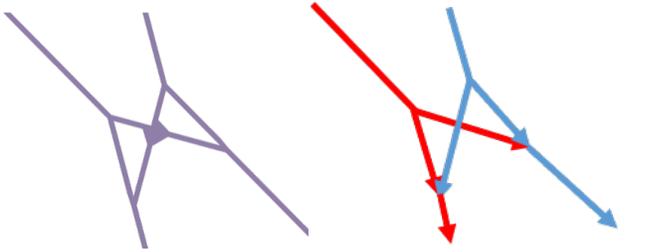 | Left: Physical geometry of a *split junction*; Right: Legal paths of the agents entering the junction. Agents entering from either the top (blue path) or the left entrance (red path) will be distributed evenly between the bottom and right exits. |
| 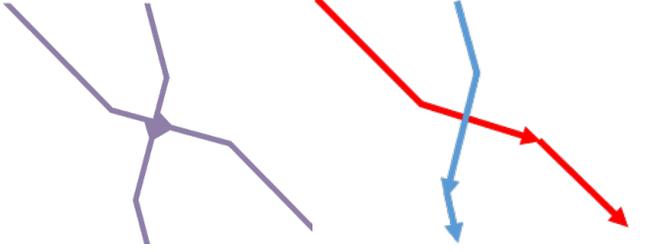 | Left: Physical geometry of a *pass junction*; Right: Legal paths of the agents entering the junction. Agents entering from either the top (blue path) or the left entrance (red path) will continue their path without making turns. |
| 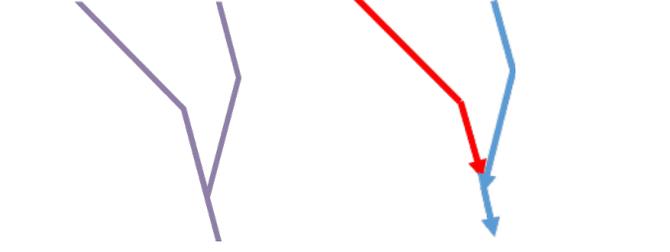 | Left: Physical geometry of a *reset-FASLE junction*; Right: Legal paths of the agents entering the junction. All agents entering from either the top (blue path) or the left entrance (red path) will be directed to the bottom exit. |
| 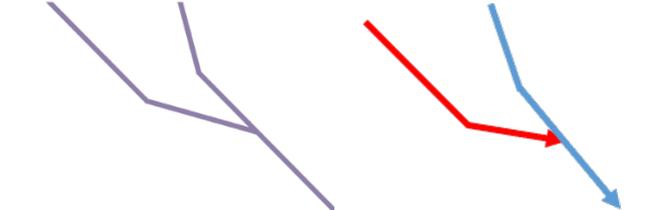 | Left: Physical geometry of a *reset-TRUE junction*; Right: Legal paths of the agents entering the junction. All agents entering from either the top (blue path) or the left entrance (red path) will be directed to the right exit. |
| 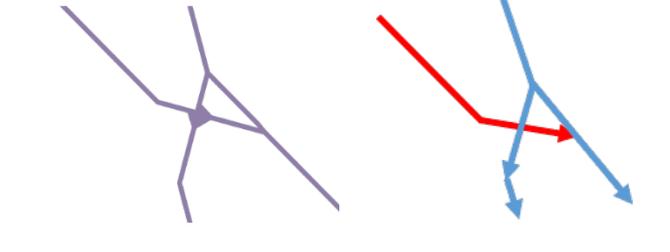 | Left: Physical geometry of a *split top junction*; Right: Legal paths of the agents entering the junction. All agents entering from the top (blue path) will be distributed evenly between the bottom and right exits. Agents entering from the left entrance (red path) will be directed to the right exit. |